\documentclass[aps,twocolumn,nofootinbib,preprintnumbers]{revtex4}
\usepackage{graphicx, float}
\usepackage{hyperref}
\usepackage{amsmath}
\usepackage{amssymb}
\usepackage[usenames,dvipsnames]{xcolor}

\def\beq{\begin{equation}}
\def\eeq#1{\label{#1}\end{equation}}
\def\eeqn{\end{equation}}
\def\beqa{\begin{eqnarray}}
\def\eeqa#1{\label{#1}\end{eqnarray}}
\def\eeqan{\end{eqnarray}}
\def\CR{\nonumber \\ }
\def\leqn#1{(\ref{#1})}

\def\Veff{V_{\rm eff}}

\def\stacksymbols #1#2#3#4{\def\theguybelow{#2}
    \def\vp{\lower#3pt}
    \def\sp{\baselineskip0pt\lineskip#4pt}
    \mathrel{\mathpalette\intermediary#1}}

\def\intermediary#1#2{\vp\vbox{\sp
     \everycr={}\tabskip0pt
     \halign{$\mathsurround0pt#1\hfil##\hfil$\crcr#2\crcr
              \theguybelow\crcr}}}

\def\gsim{\stacksymbols{>}{\sim}{2.5}{.2}}

\begin{document}

\title{Dynamics of Electroweak Phase Transition In Singlet-Scalar Extension of the Standard Model}

\author{Gowri Kurup, Maxim Perelstein}
\affiliation{Laboratory for Elementary Particle Physics, Cornell
  University, Ithaca, NY 14853, USA} 

\begin{abstract}
An addition to the Standard Model of a real, gauge-singlet scalar field, coupled via a Higgs portal interaction, can reopen the possibility of a strongly first-order electroweak phase transition (EWPT) and successful electroweak baryogenesis (EWBG). If a discrete symmetry that forbids doublet-singlet mixing is present, this model is notoriously difficult to test at the Large Hadron Collider. As a result, it emerged as a useful benchmark for evaluating the capabilities of proposed future colliders to conclusively test EWPT and EWBG. In this paper, we evaluate the bubble nucleation temperature throughout the parameter space of this model where a first-order transition is expected. We find that in large parts of this parameter space, bubbles in fact do not nucleate at any finite temperature, eliminating these models as viable EWBG scenarios. This constraint eliminates most of the region where a ``two-step" phase transition is naively predicted, while the ``one-step" transition region is largely unaffected. In addition, expanding bubble walls must not reach relativistic speeds during the transition for baryon asymmetry to be generated. We show that this condition further reduces the parameter space with viable EWBG.   
\end{abstract}

\maketitle

\section{Introduction}

In today's Universe electroweak symmetry is broken, but at very high temperatures, which prevailed immediately after the Big Bang, the symmetry was restored. The transition between the symmetric and broken phases, the Electroweak Phase Transition (EWPT), occurred when the Universe was about a nanosecond old. Understanding the nature of this transition is an interesting question in its own right. It also has profound implications for understanding the origin of matter-antimatter asymmetry in the Universe: one of the most compelling explanations of this asymmetry, the Electroweak Baryogenesis (EWBG) scenario, is only possible if the EWPT is strongly first-order~\cite{Kuzmin:1985mm}. (For a review, see~\cite{Morrissey:2012db}.)

While it is at present not possible to recreate the EWPT in the lab, it has been suggested that measurements of properties of the Higgs boson can provide indirect information about the EWPT dynamics. In the Standard Model (SM), the EWPT is an adiabatic cross-over transition~\cite{Aoki:1999fi,Csikor:1998eu,Laine:1998jb,Gurtler:1997hr}. A first-order transition is only possible in the presence of Beyond-the-SM (BSM) physics at the weak scale, with significant couplings to the Higgs sector. As a result, many models with first-order EWPT predict significant deviations of the Higgs couplings to gluons, photons, weak gauge bosons, and fermions that can already be tested at the Large Hadron Collider (LHC). In particular, supersymmetric models with stop-catalyzed first-order EWPT are already strongly disfavored~\cite{Cohen:2012zza,Curtin:2012aa,Katz:2015uja} (but not completely ruled out~\cite{Liebler:2015ddv}). A broader variety of models will be probed by increasingly precise measurements of the Higgs couplings at the LHC and the proposed $e^+e^-$ Higgs factories~\cite{Katz:2014bha,Profumo:2007wc,Damgaard:2013kva,Profumo:2014opa}. A particularly direct and powerful probe of the EWPT dynamics is provided by the Higgs cubic self-coupling, which is predicted to have significant ($\gsim 20$\% or more) deviations from the SM in most models with a first-order EWPT~\cite{Kanemura:2004ch,Noble:2007kk,Huang:2015tdv}. This prediction can be conclusively tested at the 1 TeV upgrade of the International Linear Collider (ILC)~\cite{Baer:2013cma,Fujii:2015jha,Tian:2013qmi} and the 100 TeV proton collider~\cite{Contino:2016spe}. In addition, a strongly first-order EWPT may produce a potentially observable gravitational wave signature~\cite{Grojean:2006bp}. Complementarity between collider and gravitational wave signatures has been explored in Refs.~\cite{Hashino:2016rvx,Huang:2016cjm,Hashino:2016xoj,Artymowski:2016tme,Beniwal:2017eik}. 

One of the simplest extensions of the SM in which a first-order EWPT is possible is a model with an additional gauge-singlet scalar field, $S$, coupled to the SM via a Higgs portal interaction,
\beq
V_{\rm int} \,=\, \kappa |H|^2 S^2.
\eeq{inter}    
This is the only renormalizable interaction of $S$ with the SM which is invariant under a $Z_2$ symmetry, $S\to -S$. This symmetry renders the $S$ particle stable, and it may play the role of dark matter~\cite{McDonald:1993ex,Burgess:2000yq,Feng:2014vea,Baker:2016xzo}.  It has been shown that this simple model can exhibit a strongly first-order EWPT~\cite{Espinosa:2007qk,Barger:2007im,Espinosa:2008kw,Espinosa:2011ax,Cline:2012hg,Perelstein:2016cxy,Vaskonen:2016yiu,Marzola:2017jzl}. The $Z_2$ prohibits mixing between the doublet and singlet scalars, so that the 125 GeV Higgs particle has couplings to fermions and gauge bosons that are identical to the SM. Moreover, if $m_S>m_h/2$ and the decay $h\to SS$ is kinematically forbidden, the Higgs width is also unaffected. As a result, this model presents a difficult case (sometimes dubbed a ``nightmare scenario") for tests of EWBG at future colliders, and it became an important benchmark for gauging their capabilities in this regard~\cite{Curtin:2014jma}. Some interesting recent work on this benchmark model includes suggestions for additional observables that can help cover the relevant parameter space at a 100 TeV proton collider~\cite{Curtin:2014jma,Craig:2014lda}, and an improved calculation of the thermal scalar potential~\cite{Curtin:2016urg}.

An important aspect of the EWPT in this model, which has not yet been systematically taken into account in existing studies of future collider capabilities, is the dynamics of bubble nucleation during the transition. In this paper, we aim to fill this gap. In particular, we evaluate the bubble nucleation temperature, $T_N$, throughout the parameter space relevant for EWSB and future colliders. We find that $T_N$ is often significantly lower than the critical temperature $T_c$. In fact, in large regions of the parameter space, in particular those with a ``two-step" EWPT (meaning that the $S$ field acquires a vev before the Higgs field does), we find that bubbles {\it do not nucleate} at finite temperature at all, eliminating these regions as viable EWBG scenarios. In addition, if $T_N\ll T_c$, the large difference in the vacuum energies at the stable and metastable vacua can result in ``runaway" behavior of the bubble walls, which become highly relativistic~\cite{Bodeker:2009qy}. This behavior is incompatible with the EWBG scenario. We identify the region of the parameter space (again primarily in the two-step regime) which suffers from this problem.\footnote{For recent studies of bubble-wall dynamics in this and similar models, see {\it e.g.} Refs.~\cite{Konstandin:2014zta,Kozaczuk:2015owa,Huber:2015znp,Chala:2016ykx}.} The net result of the analysis is a significant reduction of the parameter space with viable EWBG. We then comment on the implications of these additional constraints for the experimental probes of EWBG at future colliders.    

\section{Setup}

We supplement the SM with a real scalar field $S$, uncharged under any of the SM gauge groups, and impose a $Z_2$ discrete symmetry, under which $S\to -S$ and all other fields are unchanged. The tree-level scalar potential has the form
\beq
V(H; S) = -\mu^2|H|^2+\lambda |H|^4 + \frac{1}{2} m_0^2 S^2 +\frac{\eta}{4} S^4 + \kappa S^2 |H|^2,
\eeq{sc_pot} 
where $H$ is the SM Higgs doublet. If $\mu^2<0$, there is an electroweak symmetry-breaking (EWSB) minimum at zero temperature, with $\langle H\rangle = (0, v/\sqrt{2})$ and $\langle S \rangle = 0$. Depending on parameters, the potential may also have an electroweak symmetry-preserving local minimum at $\langle H\rangle = 0$ and $\langle S \rangle = s_0$. There are no stable minima with both $\langle H\rangle$ and $\langle S \rangle$ non-zero for any model parameters. The model is phenomenologically viable if the vacuum with $\langle H\rangle\not=0$ is the global minimum of the potential, $v\approx 246$ GeV and $m_h\approx 125$ GeV in this vacuum, and we restrict our attention to such parameters. This leaves three undetermined (but constrained) parameters, $m_0$, $\eta$, and $\kappa$. 

The EWPT dynamics is determined by the effective finite-temperature potential $\Veff(T)$, where $T$ is temperature. Physically, $\Veff$ is the free energy density of space filled with constant, spatially homogeneous scalar fields: 
\beq
H_{\rm bg}=(0, \frac{\varphi}{\sqrt{2}}),~~S_{\rm bg} = s~,
\eeq{eq:higgs}
and all other fields set to zero. The effective potential has the form
\beq
\Veff(\varphi, s; T) \,=\, V_0(H_{\rm bg}, S_{\rm bg}) + V_1(\varphi, s) + V_T(\varphi, s; T)\,,
\eeq{Veff}
where $V_1$ is the Coleman-Weinberg potential, and 
$V_T$ is the thermal potential~\cite{Dolan:1973qd,Weinberg:1974hy}. Both can be computed in perturbation theory. At the one-loop order,
\beqa
V_1(\varphi, s) &=& \sum_i \frac{g_i (-1)^{F_i}}{64\pi^2} \Bigl[ m_i^4(\varphi, s) \log \frac{m_i^2(\varphi, s)}{m_i^2(v, 0)} \CR & & - \frac{3}{2} m_i^4(\varphi, s) + 2 m_i^2(\varphi, s)m_i^2(v, 0)\Bigr];
\eeqa{oneloopT0}
\beqa
V_T(\varphi, s; T) &=& \sum_i \frac{g_i T^4 (-1)^{F_i}}{2\pi^2} \CR & & \hskip-2.6cm \times \int_0^\infty dx x^2 \log \left[ 1 - (-1)^{F_i} \exp \left( \sqrt{x^2 + \frac{m_i^2(\varphi, s)}{T^2}} \right) \right]
\eeqa{oneloopT} 
where the sum runs over all SM and BSM particles in the theory, and $g_i$, $F_i$ and $m_i(\varphi, s)$ are the multiplicity, fermion number, and the mass (in the presence of background fields) of the particle $i$. The counterterms included in Eq.~\leqn{oneloopT0} ensure that the tree-level Higgs mass and vev in the present, zero-temperature Universe are unchanged at one loop. The dominant contributions to $V_1$ and $V_T$ typically arise from loops of the Higgs and singlet scalar themselves. In this case, the masses $m_i(h,s)$ are obtained by diagonalizing the scalar mass matrix:
\beq
m_{1,2}^2 = \frac{1}{2}\Bigl( m_0^2 - \mu^2 + (\kappa+3\lambda) \varphi^2 + (\kappa+3\eta) s^2 \pm \Delta \Bigr),
\eeq{diag_masses} 
where $\Delta=((m_0^2 + \mu^2 + (\kappa-3\lambda) \varphi^2 + (3\eta-\kappa) s^2 )^2 + 16\kappa^2\varphi^2 s^2)^{1/2}$.
We also include contributions of the SM top quark and the electroweak gauge bosons, but ignore loops of other SM particles due to their small couplings to the Higgs. It is well known that light scalar- and gauge boson-loop contributions to $V_T$ suffer from an IR divergence. Certain classes of higher-loop contributions (so-called ``daisy diagrams") need to be resummed to obtain a good approximation for this object at $T\gg m$, where $m$ is the boson mass~\cite{Fendley:1987ef,Carrington:1991hz}. This is achieved by employing the ``ring-improved" version of $V_T$, which is obtained from Eq.~\leqn{oneloopT} by replacing the zero-temperature masses $m_i(\varphi, s)$ with thermal masses, $m^2_i\to m_i^2 + \Pi_i(T)$, where $\Pi_i$ is the one-loop two-point function at finite temperature. Recently, Ref.~\cite{Curtin:2016urg} argued that in certain regions of parameter space, further classes of diagrams may need to be resummed. We do not include these effects in the present study, leaving such improvement for future work.

At high temperature, thermal loops generate positive mass-squared for both $H$ and $S$ fields, and the energetically favored configuration has zero background fields, $(\varphi, s)=(0,0)$. As the Universe cools and thermal masses decrease, this configuration becomes unstable and the fields develop expectation values, eventually ending up in the present vacuum, $(\varphi, s)=(v, 0)$. This can occur in a number of ways. First, we distinguish between a ``one-step" transition, in which the singlet field never develops an expectation value; and a ``two-step" transition, $(0, 0)\to (0, s) \to (v, 0)$. Secondly, each transition may be first-order or second-order. In the former case, two distinct local minima of $\Veff$ coexist over a range of temperatures. At high temperatures, the ``symmetric" minimum is energetically preferred over the ``broken" minimum. (In the case of the first step of a two-step transition, ``symmetric" and ``broken" refer to vacua with $s=0$ and $s\not=0$, both of which have unbroken electroweak symmetry.) The two minima become degenerate at the critical temperature, $T_c$. As the Universe continues to cool, bubbles of the broken-minimum phase are nucleated. Nucleation probability per unit time per unit volume at temperature $T$ is given by~\cite{Dine:1992wr}
\beq
P \sim T^4 \exp (-S_3/T),
\eeq{nuc_prob}
where $S_3$ is the action of a critical bubble. We use the {\tt CosmoTransitions} code~\cite{Wainwright:2011kj} to evaluate $S_3$ numerically as a function of temperature. Nucleation temperature $T_N$ is the temperature at which the nucleation probability per Hubble volume becomes of order one; for electroweak phase transition, this corresponds to~\cite{Dine:1992wr} 
\beq
S_3/T_N\approx 100.
\eeq{NUCcrit} 
In this paper, we use this criterion to estimate $T_N$ explicitly throughout the model parameter space. (We assume $T_N=T_c$ for second-order transitions, since there is no metastable phase in that case.) Moreover, if a minimum with $s\not=0$ develops, we evaluate the nucleation temperatures for both $(0, 0)\to (0, s)$ and $(0,0)\to (v, 0)$ transitions, to determine which one occurs first.  This provides robust discrimination between one-step and two-step transitions. If the transition to EW-breaking vacuum is first-order, EWBG scenario is viable only if the baryon asymmetry created at the expanding bubble wall is not washed out by sphalerons inside the broken phase. This requires
\beq
\frac{v(T_N)}{T_N}>1,
\eeq{first}
where $v(T_N)$ is the Higgs vev at the minimum of the effective potential at the temperature $T_N$, {\it i.e.} at the time of the phase transition. There is some uncertainty as to the precise numerical criterion for baryon number preservation (see {\it e.g.}~\cite{Patel:2011th,Fuyuto:2014yia}), with $v/T$ thresholds between 0.6 and 1.4 quoted in the literature. Varying the EWBG criterion within this range has no noticeable effect on the conclusions of our study, such as the phase diagrams presented below. 

Another potentially important aspect of a first-order EWPT is the velocity of the expanding bubble wall. The wall experiences outward pressure due to the difference in energy densities of the symmetric and broken vacua, $V_{\rm vac}({\rm sym})-V_{\rm vac}({\rm br})$, where $V_{\rm vac}=V_0+V_1$. It also experiences pressure $P$ from the thermal plasma of particles that it is moving through; since the particles are heavier in the broken phase than in the symmetric one, the effect of this pressure is to slow the wall down. The balance between these two forces determines whether the wall reaches a non-relativistic terminal velocity, or continues to accelerate until it becomes highly relativistic. In the latter case, electroweak baryogenesis cannot occur, since there is not enough time to generate the baryon-antibaryon asymmetry in the region in front of the advancing bubble wall. Thus, to find viable models of EWBG one must not only require a strongly first-order EWPT, but also demand that the bubble wall does not reach $v_{\rm wall}\sim 1$~\cite{Bodeker:2009qy}. Relativistic wall motion occurs if 
\beq
V_{\rm vac}({\rm sym})-V_{\rm vac}({\rm br})-P>0,
\eeq{BM_crit}  
where the pressure $P$ is calculated assuming $v_{\rm wall}\sim 1$. This calculation was performed by Boedeker and Moore in Ref.~\cite{Bodeker:2009qy}, with the result
\beq
P \approx \sum_i \left( m_i^2({\rm br})-m_i^2({\rm sym})\right) \frac{g_i T_N^2}{4\pi^2}  \tilde{J}_i \left(\frac{m_i^2({\rm sym})}{T_N^2}\right),
\eeq{P_BM}
where 
\beq
\tilde{J}_i(x) = \int_0^\infty \frac{y^2 dy}{\sqrt{y^2+x}} \frac{1}{e^{\sqrt{y^2+x}}+(-1)^{F_i}}.
\eeq{Jtilde}
We will apply the {\it Bodeker-Moore (BM) criterion}, Eq.~\leqn{BM_crit}, to further constrain the viable parameter space for EWBG. Note that Ref.~\cite{Bodeker:2009qy} argued that if the BM criterion is satisfied, the walls will exhibit ``runaway" behavior, continuing to accelerate indefinitely once they are relativistic. Very recently, the analysis has been refined to include the effect of transition radiation by charged particles crossing the bubble wall, with the result that the wall velocities are limited~\cite{Bodeker:2017cim}. However, the newly established speed limit, $\gamma\sim 1/\alpha$, is still highly relativistic, so that the conclusions regarding viability of EWBG are unaffected.   

\section{Results}

\begin{figure}[t!]
\centering
\includegraphics[width=.45\textwidth]{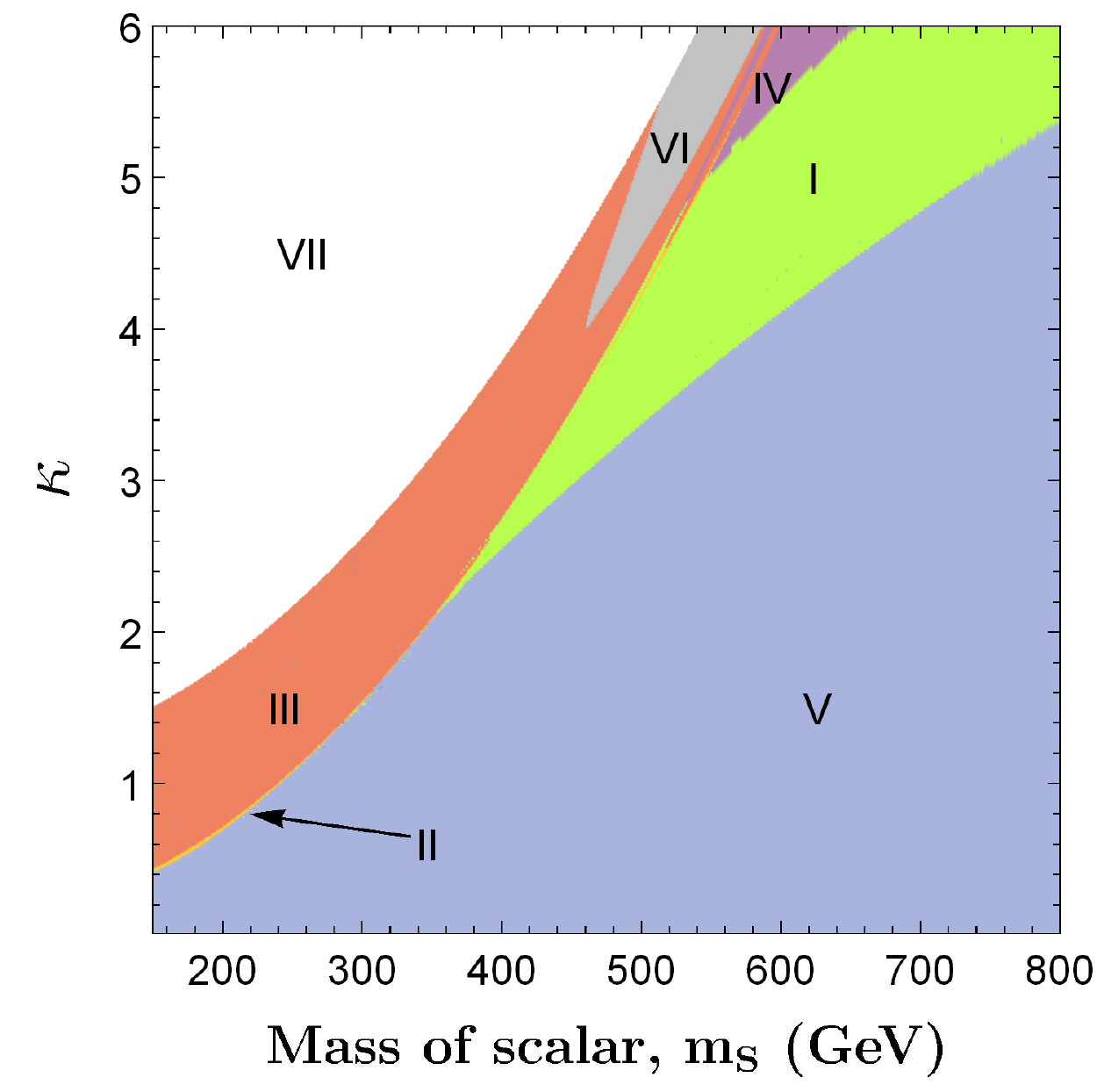}
\caption{Phase transition dynamics in the $m_S-\kappa$ plane, with $\eta=\eta_{\rm min}+0.1$. Region I (green): one-step strongly first-order transition; Region II (yellow): two-step transition with strongly first-order electroweak-symmetry breaking step; Region III (red): no thermal phase transition (a would-be two-step transition, but bubbles fail to nucleate); Region IV (purple): same as red, with a would-be one-step transition; Region V (blue): second-order transition; Region VI (gray): no viable EWSB at zero temperature; Region VII (white): non-perturbative regime ($\eta>10$).}
\label{fig:massscans1}
\end{figure}

\begin{figure}[t]
\centering
\includegraphics[width=.45\textwidth]{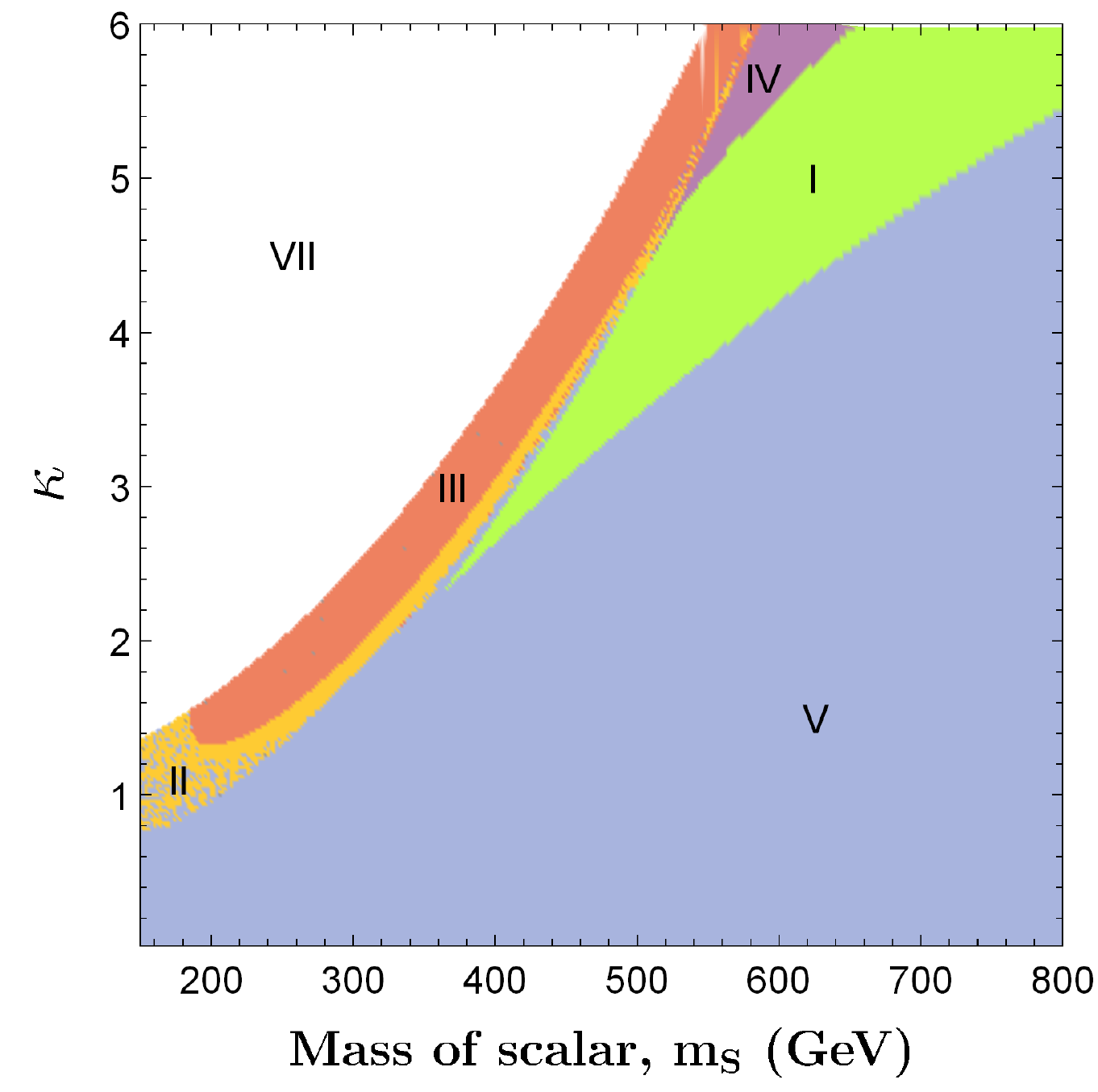}
\caption{Phase transition dynamics in the $m_S-\kappa$ plane, with $\eta=\eta_{\rm min}+2.5$. Same labeling and color code as in Fig.~\ref{fig:massscans1}.}
\label{fig:massscans2}
\end{figure}

\begin{figure}[t]
\centering
\includegraphics[width=.45\textwidth]{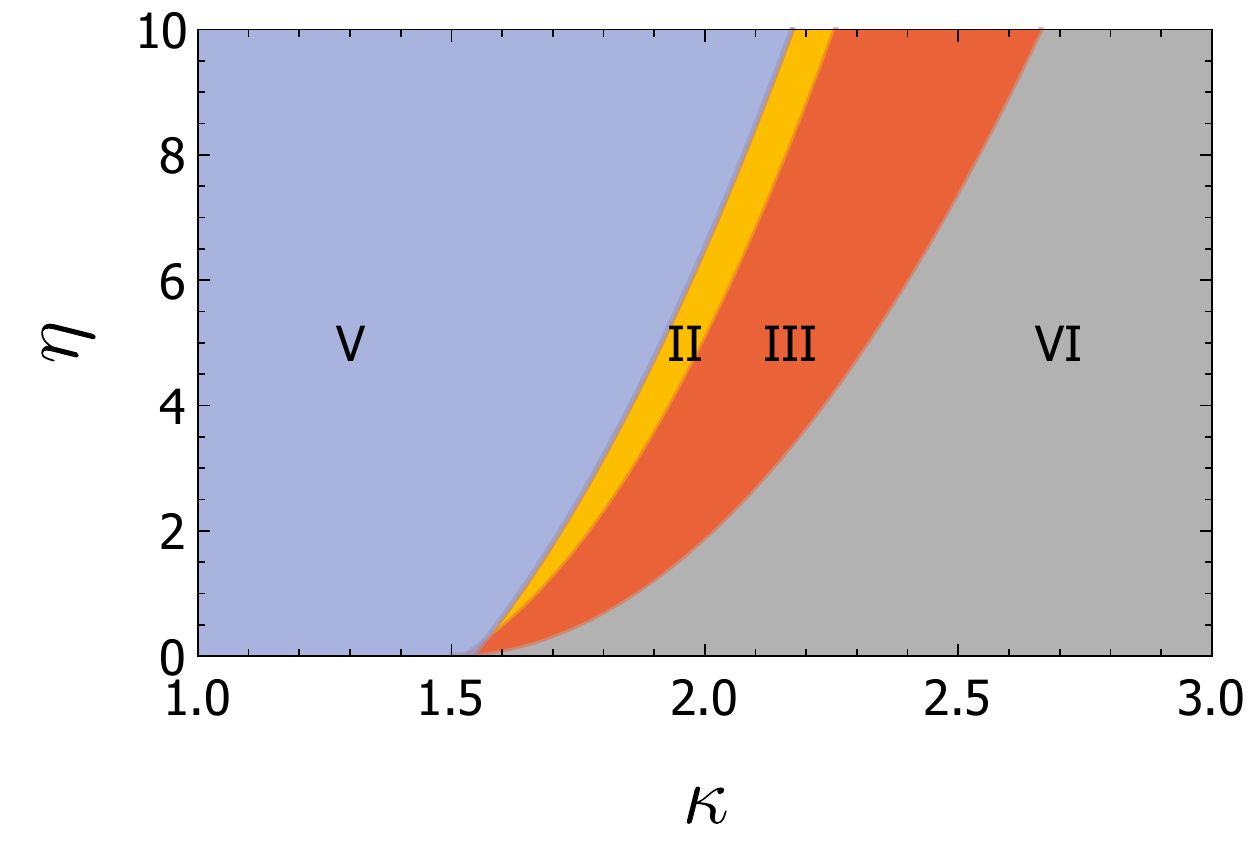}
\caption{Phase transition dynamics in the $\kappa-\eta$ plane, with $m_S=300$ GeV. Same labeling and color code as in Fig.~\ref{fig:massscans1}.}
\label{fig:etascans}
\end{figure}

\begin{figure}[t!]
\centering
\includegraphics[width=.45\textwidth]{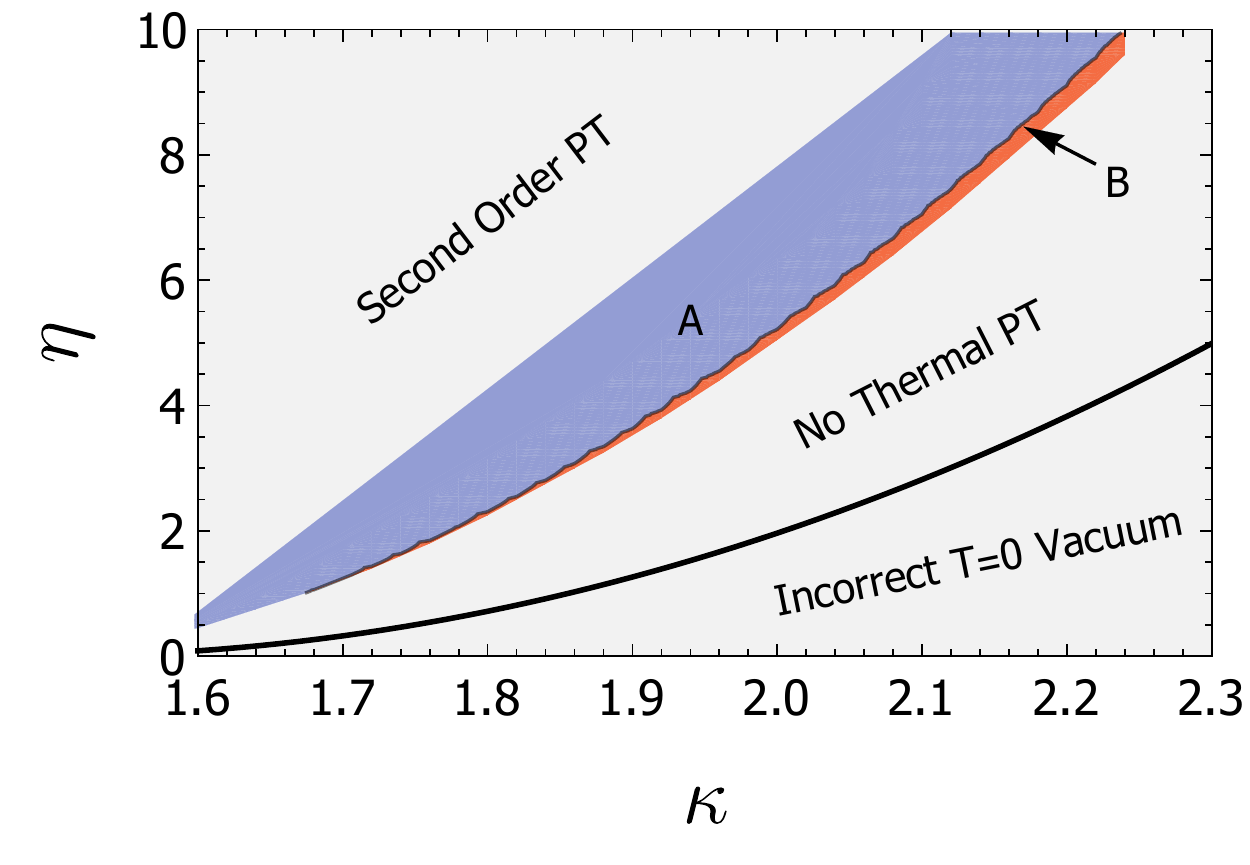}
\caption{Phase transition dynamics in the $\kappa-\eta$ plane, with $m_S=300$ GeV. In region B (red) bubble walls accelerate to relativistic speeds and EWBG cannot occur, while in region A (blue) EWBG is possible.}
\label{fig:BMexclusion}
\end{figure}

We performed a comprehensive scan of the model parameter space, $(m_0, \kappa, \eta)$. For each point in the scan with viable zero-temperature vacuum structure, we determine the transition history (one-step or two-step); critical temperature and transition order (for each step, in the case of two-step transition); nucleation temperature, for each first-order transition; and, in the case of first-order EWSB transition, whether or not the BM criterion is satisfied. The results are summarized in a series of two-dimensional slices through the parameter space, Figs.~\ref{fig:massscans1},~\ref{fig:massscans2},~\ref{fig:etascans},~\ref{fig:BMexclusion}. For clarity, we trade the scalar mass parameter $m_0$ for the physical mass of the singlet scalar, $m_S=\left(\partial^2 V_{\rm vac}(v, 0)/\partial S^2\right)^{1/2}$, in these plots.

The main new result is that in large parts of the parameter space where a naive criterion used in previous studies suggests a strongly first-order electroweak phase transition, bubble nucleation in fact does not occur at any finite temperature, so there is no thermal phase transition at all. Instead, the system becomes trapped in the metastable state with unbroken electroweak symmetry, either at the origin or at $(0, s_0)$. Eventually, it may transition to the stable EW-breaking vacuum by tunneling at $T=0$, and such models may be viable descriptions of today's Universe; however, they do not provide viable scenarios for electroweak baryogenesis. Any discussion of collider experiments required to test EWBG must take this constraint into account. 

\begin{figure}[th]
\centering
\includegraphics[width=.45\textwidth]{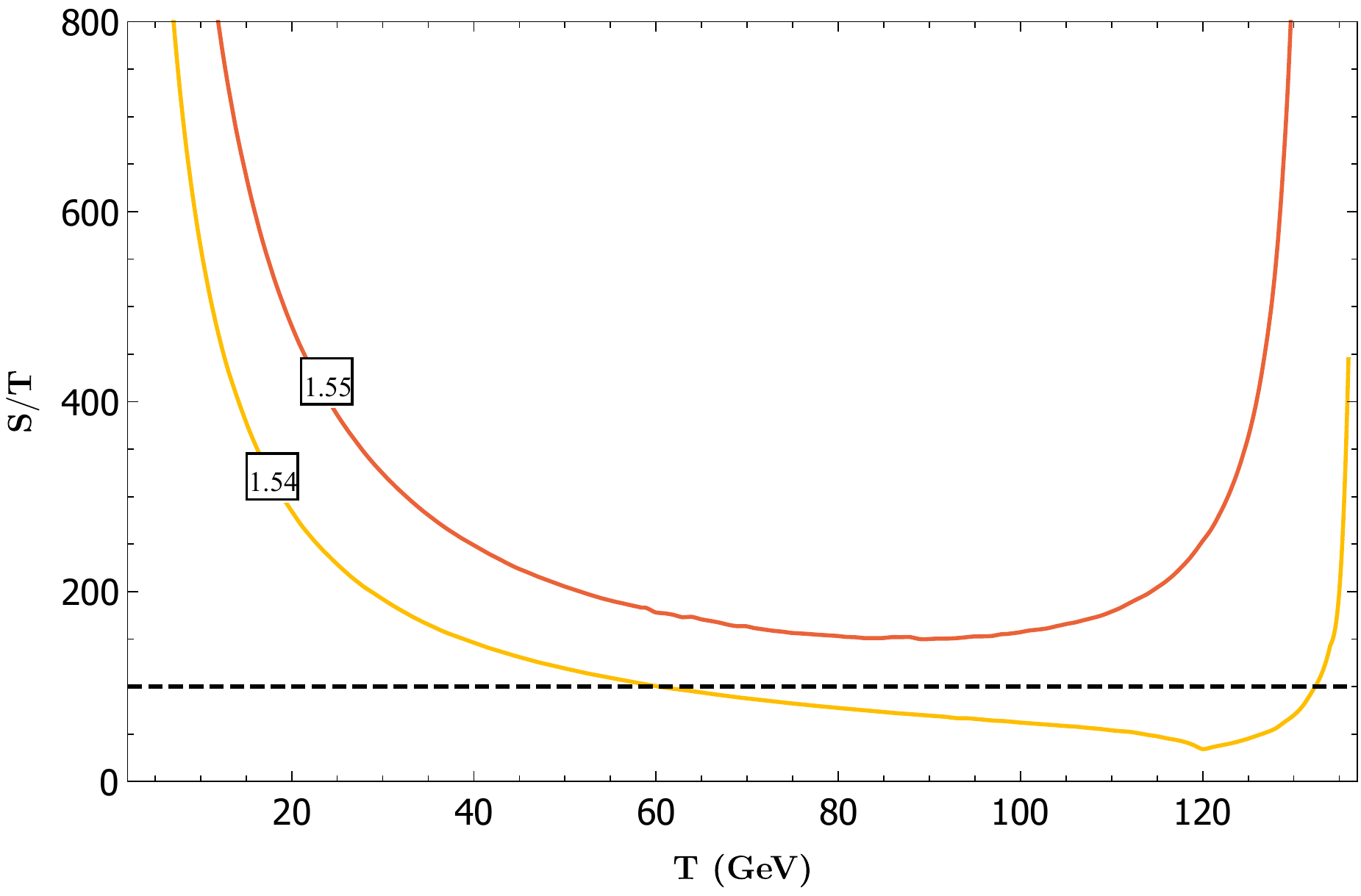}
\caption{Ratio $S_3/T$, where $S_3$ is the critical bubble action, for $m_S=300$ GeV and $\kappa=1.55$ (red) and 1.54 (yellow). For both points, a two-step first-order transition is naively expected. In fact, thermal transition does not occur at $\kappa=1.55$.}
\label{fig:SoverT}
\end{figure}

\begin{figure}[t]
\centering
\includegraphics[width=.45\textwidth]{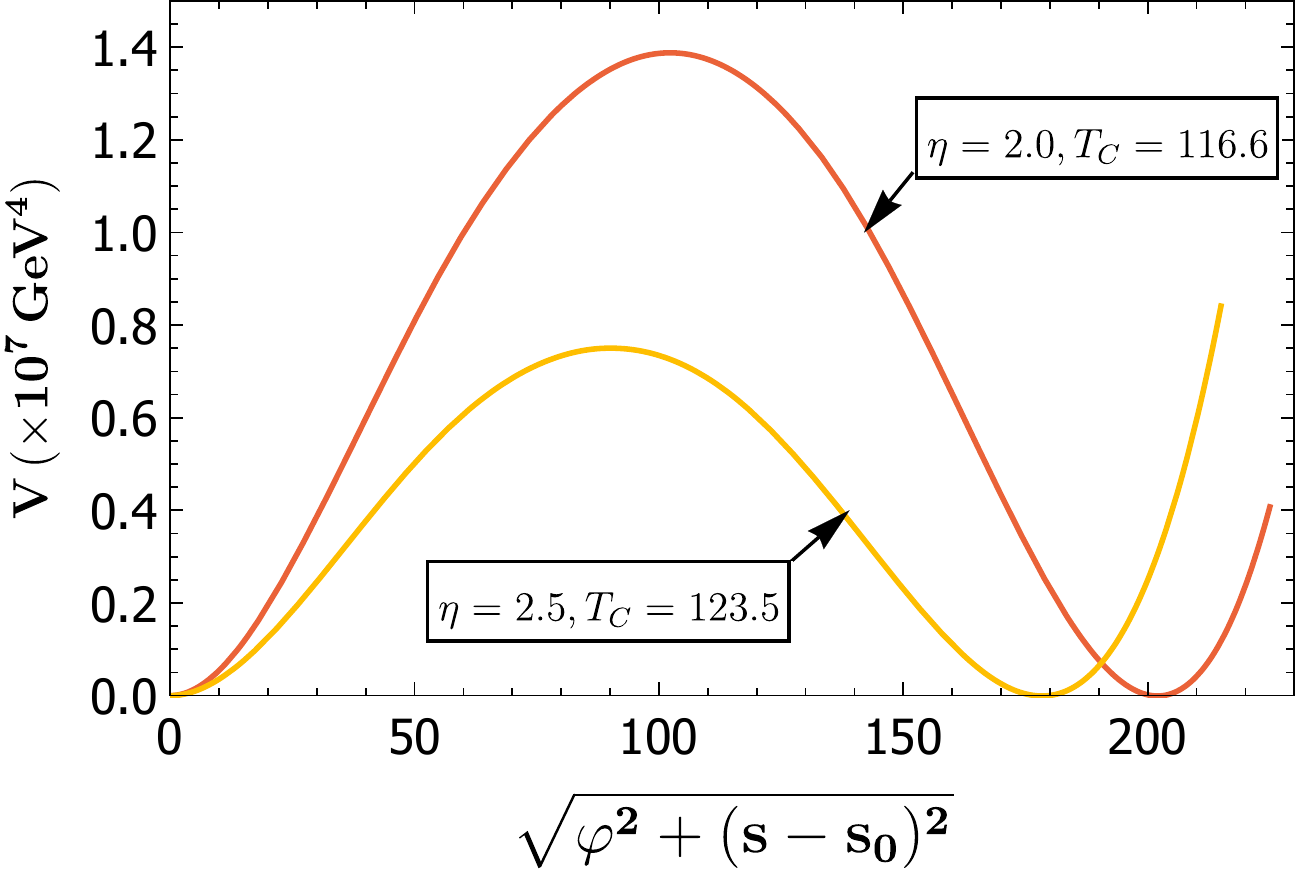}
\caption{Thermal potential at the critical temperature, along the line in field space connecting the EW-symmetric and broken vacua, for $m_S=300$ GeV, $\kappa=1.8$, and two representative values of $\eta$, 2.0 (red) and 2.5 (yellow). For both points, a two-step first-order transition is naively expected. In fact, thermal transition does not occur at $\eta=2.0$.}
\label{fig:eta_pot}
\end{figure}

A striking example is provided by Fig.~\ref{fig:massscans1}. Following Ref.~\cite{Curtin:2014jma}, in this plot we fixed the singlet quartic coupling at $\eta=\eta_{\rm min}+0.1$, where $\eta_{\rm min}=\lambda m_0^4/\mu^4$ is the minimum value for which $(v, 0)$ is the global minimum of the tree-level potential. Essentially the entire region where a two-step transition would be expected is eliminated due to failure to nucleate bubbles at any temperature. A two-step thermal phase transition can only occur in a very narrow sliver of parameter space at the bottom of this region, shown in yellow in Fig.~\ref{fig:massscans1}. The reason is that in the two-step region, a large potential barrier between the EW-preserving and EW-breaking vacua is present at any temperature, down to $T=0$. As a result, the critical bubble action $S_3$ is limited from below, and if this limit is sufficiently large, the bubble-nucleation criterion~\leqn{NUCcrit} is never satisfied. This is illustrated in Fig.~\ref{fig:SoverT}.\footnote{There is some uncertainty as to the precise numerical value of the right-hand side in Eq.~\leqn{NUCcrit}. We use 100 in Figs.~\ref{fig:massscans1}-\ref{fig:BMexclusion}. We have checked that varying this threshold by 20\% does not have a significant effect on the phase diagrams. The reason is clear from Fig.~\ref{fig:SoverT}: at the boundary between the regions with and without thermal phase transition, small changes in model parameters lead to large changes in the critical bubble action.} In contrast, in the one-step region, there is no EW-preserving vacuum at $T=0$ at tree level. This guarantees bubble nucleation at finite temperature, unless the couplings are very strong and loop corrections become important. Consequently, most of the one-step region survives this constraint. 

The shape of the potential, and hence dynamics of bubble nucleation, depend on the singlet quartic coupling $\eta$ as well as $m_S$ and $\kappa$. We find that for larger $\eta$, it is easier to find points in the two-step region where the thermal EWPT does occur, and is strongly first-order. The reason is that as $\eta$ is increased, the critical temperature of the transition between the EW-symmetric and broken vacua increases, and both the height and the width of the potential barrier decrease; see Fig.~\ref{fig:eta_pot}. This makes tunneling between the two vacua easier, allowing a thermal phase transition to occur. The effect of varying $\eta$ on the viable parameter space is illustrated in Figs.~\ref{fig:massscans2} and~\ref{fig:etascans}. Note, however, that even at large $\eta$, most of the two-step region is eliminated by the requirement of bubble nucleation at non-zero temperature. 

Even if this requirement is satisfied, models in which the nucleation temperature $T_N$ significantly below the critical temperature $T_c$ are likely to fail the BM criterion for relativistic bubble wall motion. This is because in this case, the symmetry-breaking vacuum would typically have a significantly lower vacuum energy at $T_N$ compared to the symmetric vacuum, resulting in a strong outward pressure on the bubble wall. To check this, we implemented the BM criterion, Eq.~\leqn{BM_crit}, in our scans. The result, shown in Fig.~\ref{fig:BMexclusion}, is consistent with expectations. The BM criterion eliminates a region bordering that where no thermal EWPT occurs, since by continuity this is the region where $T_N$ is the lowest. This extra constraint must also be taken into account in the discussion of collider probes of EWBG. 

\section{Discussion}

We re-considered the dynamics of EWPT in a model with a singlet scalar field $S$ coupled to the SM via a $Z_2$-symmetric Higgs portal, Eq.~\leqn{inter}. We found that the requirements of thermal EWPT (bubble nucleation at non-zero temperature) and non-relativistic bubble wall motion eliminate much of the parameter space that was previously thought to provide viable EWBG models. In particular, most of the parameter space where a two-step phase transition was thought to occur, is now eliminated. The effect of the new requirements in the region where a one-step transition was expected is less significant.   

The model studied here has recently emerged as a useful benchmark for planning the physics program at future colliders. While absence of mixing between doublet and singlet states makes this model challenging to probe at the LHC, Ref.~\cite{Curtin:2014jma} argued that the proposed future facilities will be able to probe the EWBG scenario in this model conclusively. This can be achieved with a combination of Higgs cubic coupling measurements~\cite{Noble:2007kk}, direct Higgs portal searches in channels such as $pp\to VSS, qqSS$~\cite{Curtin:2014jma,Craig:2014lda}, and a very precise measurement of $\sigma(e^+e^-\to Zh)$ at electron-positron Higgs factories~\cite{Craig:2013xia,Katz:2014bha,Craig:2014una}. The new constraints considered here reduce the parameter space with viable EWBG, which should in principle make the colliders' task easier. However, comparing the predictions for collider observables in Ref.~\cite{Curtin:2014jma} with the new constraints presented here indicates that the newly eliminated parts of the parameter space are the ones with the {\it strongest} collider signals. This should not be surprising, since for a given $m_S$, our constraints place an upper bound on $\kappa$, while all collider observables deviate from the SM more with growing $\kappa$. (Note that $T=0$ collider observables are independent of $\eta$ up to the one-loop order, since $\eta$ does not enter $V_{\rm vac}$ in the present vacuum at one loop). Thus, the sensitivity goals established by previous studies as benchmarks for future colliders remain unchanged.  

Our findings seem to indicate that in this model, a viable two-step first-order transition occurs only in a rather special, narrow region of the parameter space, in effect requiring some degree of tuning between the model parameters. This may appear to make this scenario ``unlikely". However, it is important to remember that the parameters of the model may emerge from a more fundamental theory at higher energy scales, which may in fact correlate parameters that we treat as independent. Therefore, it would be incorrect to interpret our results as in any way reducing the motivation for an experimental program that will address the viability of EWBG in this model.

\acknowledgments{We are grateful to Andrey Katz for discussions which initiated this investigation; to Nicolas Rey-Le Lorier for collaboration at the early stages of the project; to Carroll Wainwright for valuable communications regarding {\tt CosmoTransitions}; to Yu-Dai Tsai for helpful discussions; and to Michael Baker for pointing out a typographical error in the original version of this preprint. This work is supported by the U.S. National Science Foundation grant PHY-1316222, and by the Simons Foundation grant \#399528.}

\bibliography{refs}
\bibliographystyle{apsper}

\end{document}